\documentclass{elsart}
\usepackage{graphics}
\begin{document}
\begin{frontmatter}
\textbf{Preprint MPI~H-V2-2001, updated}\\
\title{On the form factor of physical mesons\\ 
       and their distribution function}
%
%
\author{Hans-Christian Pauli}
\address{Max-Planck Institut f\"ur Kernphysik, D-69029 Heidelberg, Germany}
\date{31 July 2001}
\begin{abstract}
    This work addresses more to the technical rather than to the 
    physical problem, how to calculate analytically
    the form factor $F(Q)$, 
    the associated mean-square radius $<r^2>$,
    and the distribution function $\Phi(x,Q^2)$
    for a given light-cone wave function
    $\Psi_{q\bar q}(x,\vec k _{\!\perp})$ of the pion.
    They turn out to be functions of only one dimensionless
    parameter, which is the ratio of the constituent
    quark mass and an effective Bohr momentum which
    measures the width of the wave function in momentum space.
    Both parameters are subject to change in the future, 
    when the presently used solution for 
    the over simplified $\uparrow\downarrow$-model 
    will be replaced by something better.   
    Their relation to and agreement with experiment is 
    discussed in detail.~---
    The procedure can be generalized also to other hadrons. 
\end{abstract}
\maketitle
\end{frontmatter}
%
\section{Introduction and Motivation}
\label{sec:1}
A quantitative measure of hadronic sizes is the mean-square radius.
Its experimental value for the pion ($\pi^+$) is \cite{am84}
$\sqrt{\langle r^2\rangle }= 0.67\pm 0.02$~fm. 
One determines it by first measuring the electro-magnetic 
form factor $F(Q^2)$ for sufficiently small values of the 
(Feynman-four-) momentum transfer $Q^2=-(p_{e}-p_{e'})^2$, 
and then taking the derivative at sufficiently small $Q^2$, 
{\it i.e.}
\begin{equation}
   \langle r^2\rangle  = 
   \left.-6\ \frac{dF(Q^2)}{dQ^2}\right\vert_{Q^2=0} 
.\label{eq:1}\end{equation}
The electro-magnetic form factor can also be calculated.
One of the most remarkable simplicities 
of the light-cone formalism \cite{BroPauPin98}
is that one can write down an exact expression.  
As was first shown by Drell and Yan \cite{dry70},  
it is advantageous to  choose a special coordinate frame 
to compute form factors and other current matrix elements
at space-like photon momentum. 
In the Drell frame \cite{leb80}, the four-momentum transfer is
$ -q_\mu q^\mu \equiv Q^2 = \vec q _{\!\perp} ^{\;2}$.   
The space-like form factor for a hadron 
is just a sum of overlap integrals analogous to 
the corresponding non-relativistic formula \cite{dry70}. 
The general formula in \cite{BroPauPin98} holds for any composite hadron 
and any initial or final spins $S$,
but is particularly simple for a spin-zero hadron like a pion. 
It works with the wave functions 
$\Psi_{n}=\Psi_{q\bar q},\Psi_{q\bar q g},\dots$, 
which are the Fock-space projections of the hadrons eigenstate. 
The total wave function for a meson is for example 
$ \vert\Psi_{meson}\rangle = \sum_{i} ( 
   \Psi_{q\bar q}(x_i,\vec k_{\!\perp_i},\lambda_i)  \vert q\bar q\rangle   +
   \Psi_{q\bar q g}(x_i,\vec k_{\!\perp_i},\lambda_i)\vert q\bar q g\rangle +
   \dots)$. 
The computation of these projections 
is the goal of the light-cone approach 
to the bound-state problem in gauge theory \cite{BroPauPin98}, 
by solving $H_{LC}\vert\Psi\rangle = M^2\vert\Psi\rangle$,
with the eigenvalues $M^2$ being the invariant mass-squares 
of the physical mesons.

\begin{figure}
\begin{minipage}[t]{67mm}
  \resizebox{0.95\textwidth}{!}{\includegraphics{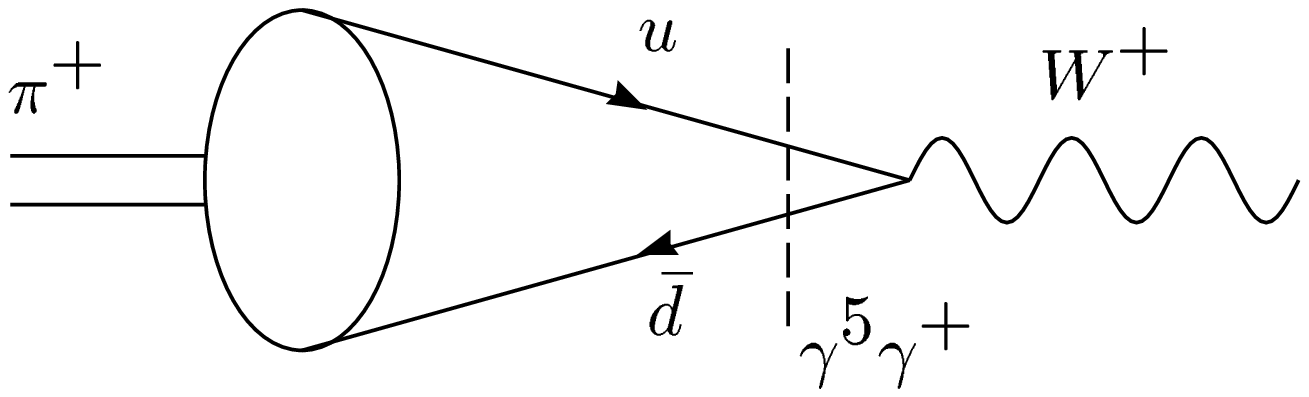}}
  \caption{\label{fig:2}  
   Matrix element of the charged axial-vector current
   controlling the decay $\pi\rightarrow e\bar\nu$.}
\end{minipage}
\ \hfill
\begin{minipage}[t]{67mm}
  \resizebox{0.99\textwidth}{!}{\includegraphics{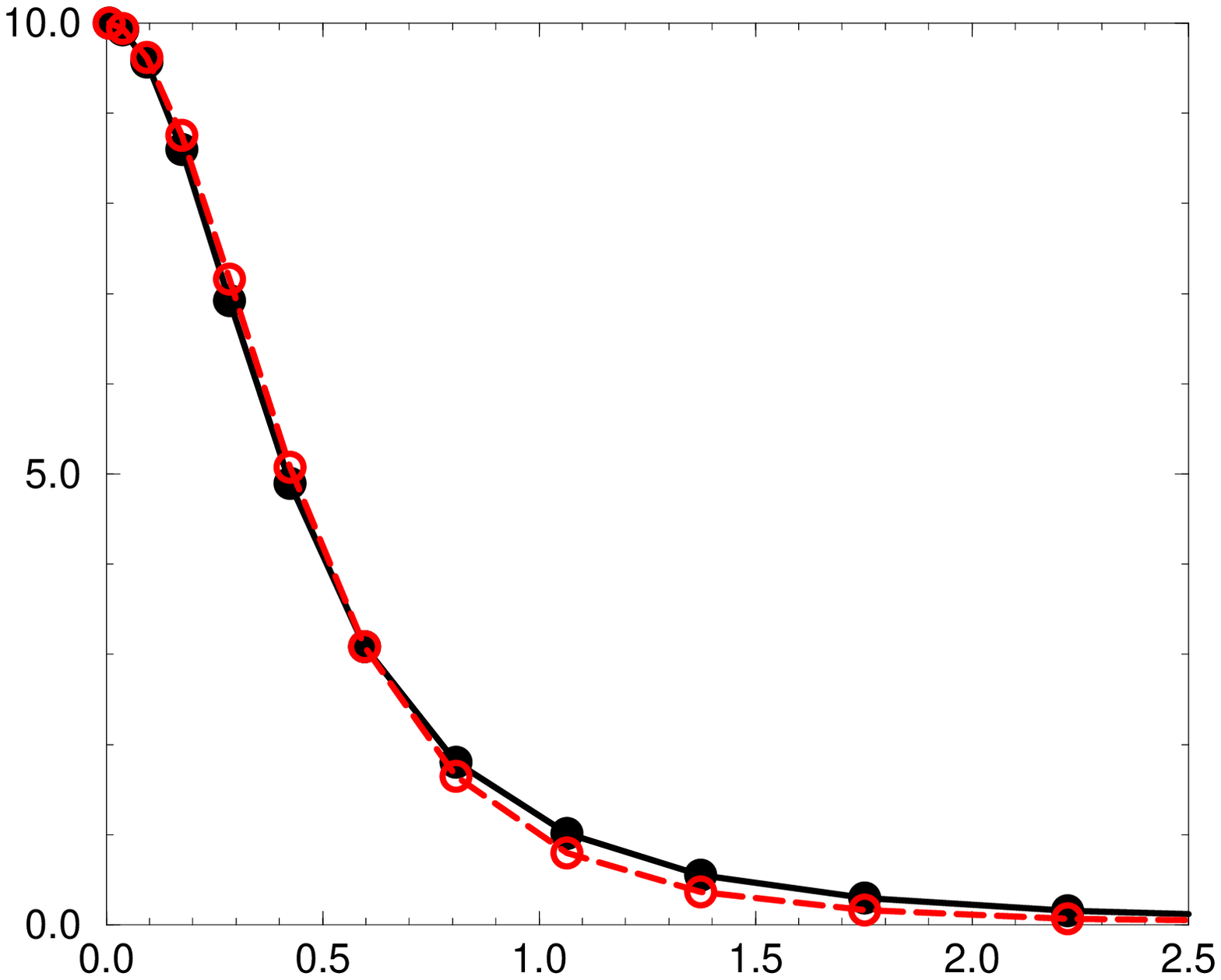}}
  \caption{\label{fig:w2}   
   The reduced wave function $\varphi(p)$ 
   for the pion is plotted versus $p$
   in an arbitrary normalization. 
   The filled circles indicate the numerical results,
   the open circles the fit to Eq.(\protect{\ref{eq:ff}}).
}\end{minipage} 
\end{figure}

It is advantageous to know $Z_2$, 
the probability amplitude for finding the two-particle Fock-state
$\Psi_{u\bar d}$ in the pion total eigenstate 
$\vert\Psi_{pion}\rangle$,
see {\it f.e.} \cite{BroPauPin98}.
It can be obtained by computing  
the leptonic decay of the $\pi^+$ \cite{leb80,BroPauPin98} 
as shown in Fig.~\ref{fig:2},
\begin{equation}
  Z_2\int \frac{dx\,d^2\vec k_{\!\perp}}{N_{cn}}\, 
  \psi(x,\vec k_{\!\perp})
  = {f_\pi \over 2\sqrt{n_c}}, \quad
  N_{cn}=\cases{\phantom{\sqrt{}}
        16\pi^3 ,& in \cite{leb80};\cr
  \sqrt{16\pi^3},& present.\cr}
\label{eq:Z2}\end{equation}
The factor $N_{cn}$ depends on the continuum normalization 
of the pion wave function, and $n_c=3$ is the number of colors.
In Brodsky's convention, the empirical pion decay constant \cite{dum83} 
is $f_\pi \sim 130\mbox{ MeV}/\sqrt{2}\approx \hbox{93~MeV}$. 
The relation involves only the $L_z=S_z=0$ component of the
general $u\bar d$ wave function, where
$\psi(x,\vec k_{\!\perp})\equiv
 \Psi_{u\bar d}(x,\vec k_{\!\perp};\uparrow\downarrow)/Z_2$
is the (normalized) probability amplitude for finding 
the quarks with anti-parallel helicities, particularly
for finding the up-quark with 
longitudinal momentum fraction $x$ and 
transversal  momentum $\vec k_{\!\perp}$,
and the down antiquark with $1-x$ and $-\vec k_{\!\perp}$.
Their respective charges are $e_1$ and $e_2$, respectively, 
with $e_1+e_2=1$.

The analytical calculation of $Z_2$ is the second aim of this work.
As shown below, $Z_2$ is not only finite but even large, 
see for example \cite{BHL80,BHL81}.
The third aim of this work is to calculate
the pions distribution function
\begin{equation}
  \Phi_{u\bar d}(x,Q^2)=\pi\int_0^{Q^2} dk^2_{\!\perp}
  \ \Psi_{u\bar d}(x,k_{\!\perp};\uparrow\downarrow)
,\label{eq:Phi}\end{equation}
where the transversal momenta are integrated up to some momentum 
scale $Q^2$ \cite{BroPauPin98,leb80}.
The distribution function continues to play an important role 
and we cite only a small fraction of the available literature
\cite{leb80,EfR80,BBGG81,BrL81,Bro82,ChZ84,BrM88,FMS93,JaK93}.

It appears as if this paper is loaded with formalism, but
an attempt was made to proceed pedagogically.
In section~\ref{sec:2}, we expand on general considerations 
and give the explicit formulas for the purpose 
of definition and notation.
In section~\ref{sec:3}, 
the same results as in a precursor to this work \cite{PaM01}
are derived in a simpler way.
The non-interested reader may proceed to 
section~\ref{sec:4}, where the necessary integrations
of Eqs.(\ref{eq:1}-\ref{eq:Phi}) are carried out
explicitly quoting only the definitions and the results.
All intermediate steps are left out, albeit they contain
the lion's share of this work.
The emphasis of the present work is on the analytical evaluation
of expressions taken from the literature, but 
some physical aspects are discussed in section~\ref{sec:5}.
In section~\ref{sec:6}, explicit expressions of the distribution 
function are discussed, and finally, 
in section~\ref{sec:7}, the necessarily compact presentation 
of this work is summarized.
%
%
\section{General considerations}
\label{sec:2}
It is known empirically, that the form factor at low $Q^2$ has 
essentially mono-pole structure \cite{am84}, and that
the mean-square radius is essentially all the 
information there is. 
In the sequel, we restrict considerations
to the contribution of the 2-particle Fock-state to the
form factor of the pion, {\it i.e.} to 
\begin{equation} 
   F_2(Q) =
   \!\!\int\!\!dx\,d^2\vec k_{\!\perp}\!\!\left(
   e_1\psi(x,\vec k_{\!\perp}+(1-x)\vec q _{\!\perp})+
   e_2\psi(x,\vec k_{\!\perp}-  x  \vec q _{\!\perp}) \right)
   \psi(x,\vec k_{\!\perp}) 
.\label{eq:4}\end{equation}
Since $\psi(x,\vec k_{\!\perp})$ is normalized, $F_2(0) = 1$. 
The associated mean-square radius, the `size', is defined by 
\begin{equation}
   R_2^2 \equiv \langle r^2\rangle_2  = 
   \left.-6\ \frac{dF_2(Q^2)}{dQ^2}\right\vert_{Q^2=0} 
.\end{equation}
The three quantities $R_2$, $Z_2$ 
and $\Phi_{u\bar d}(x,Q^2)$ are of physical interest, 
since they can be measured \cite{am84,dum83,Ash00},
to some extent, see also section~\ref{sec:5}.
For a given light-cone wave function $\psi(x,\vec k_{\!\perp})$
their calculation is straight-forward.
But here is a problem:
The bound-state equation for non-perturbative QCD, 
which should define this function, continues to be a 
challenge \cite{BroPauPin98}.
Recently, however, some progress was made with the
oversimplified $\uparrow\downarrow$-model \cite{Pau99b}
for the pion and other pseudo-scalar mesons.
It produces a numerical $\psi(x,\vec k_{\!\perp})$.
But the three-dimensional numerical integration required in 
Eq.(\ref{eq:4}) and its subsequent derivation with respect to $Q^2$ 
is cumbersome and may be numerically inaccurate.
It might be (numerically) more reliable to suitably
parametrize $\psi(x,\vec k_{\!\perp})$ 
and to perform the quadratures
invoked by Eq.(\ref{eq:4}) analytically.
In fact, a certain form of the parametrization is imposed 
by the structure of the bound-state equation in general,
see however also the precautions mentioned in Section~\ref{sec:7}.

Quite in general, one is able to write down an integral
equation in the three variables $x$ and $\vec k _{\!\perp}$ for 
the wavefunction $\psi(x,\vec k _{\!\perp})$,
see {\it f.e.} \cite{BroPauPin98,Pau99b}.
The solution of such an equation is numerically non-trivial,
among other reasons, because the longitudinal momentum fractions 
are limited to $0\leq x \leq 1$.
It is therefore advantageous to substitute the integration variable $x$ 
by an other integration variable $k_z$, 
$-\infty\leq k_z\leq \infty$,  
which has the same range than either of the two transversal momenta
$\vec k _{\!\perp}$.  
For equal quark masses $m_u=m_d=m$, 
the substitution by the generalized Sawicki transform is simple, 
see for example \cite{Pau00d}, {\it i.e.}
\begin{equation}
   x(k_z) = \frac{1}{2}\left(1 +
   \frac{k_z}{\sqrt{m^2 + \vec k_{\!\perp}^{\,2} + k_z^2}}\right)
\quad\Longleftrightarrow\quad
   k_z^2 = (m^2+\vec k_{\!\perp}^{\,2}) 
   \frac{(x-\frac{1}{2})^2}{x(1-x)}
.\label{eq:xkz}\end{equation}
Formally, the three integration variables $k_z$ and $k_{\!\perp}$
look like a conventional 3-vector $\vec p \equiv (k_z,\vec k_{\!\perp})$. 
The substitution and the associated Jacobian seems to destroy
the hermitian property of the kernel, but only superficially, 
since it can be restored by substituting
\begin{equation}
   \psi(x,\vec k _{\!\perp}) = \varphi(k_z,\vec k _{\!\perp})
   \ \frac{\sqrt{1+\vec p ^{\;2}/m^2}}{\sqrt{x(1-x)}} 
.\label{eq:psi}\end{equation}
Mathematically, the so obtained integral equation
for the reduced wave function $\varphi(\vec p)$
is identical with the original integral equation 
for the light-cone wave function $\psi(x,\vec k _{\!\perp})$.
It looks like an integral equation in usual momentum space ($\vec p$),
but continues to be a relativistically correct front-form equation.
The reduced wave function $\varphi(\vec p)$
describes a bound state and must decrease for $p\rightarrow\infty$
with a certain scale $p_a$, either exponentially or like
a power. As discussed in \cite{PaM01} a power law is more likely.
In the present context it can suitably be parametrized by 
\begin{equation}
   \varphi(p) = \frac{\mathcal{N}}{\left(1+{p^2}/{p_a^2}\right)^{2}}
.\label{eq:ff}\end{equation}
The analogue to the Bohr momentum $p_a$ can be fitted 
to the numerical solution.
An example is given in Fig.~\ref{fig:w2}.
More details can be found in \cite{BroPauPin98,Pau00b,Pau99b}.

To be even more specific and concrete, 
the integral equation in the  $\uparrow\downarrow$-model 
is taken from \cite{Pau00b} as an illustration without proof, 
{\it i.e.}
\begin{eqnarray}
   M^2\varphi(\vec p) &=& \left[4m^2+4\vec p^{\,2}\right]
   \varphi(\vec p)
\nonumber\\ &-& 
   \frac{4}{3}\ \frac{\alpha}{2\pi^2}\int\frac{d^3\vec p\,'}{m} 
   \left(\frac{4m^2}{(\vec p - \vec p\,')^{\,2}} + 
   \frac{2\mu^2}{\mu^2+(\vec p - \vec p\,')^{2}} \right)
   \varphi(\vec p\,')
.\label{eq:bsEq}\end{eqnarray}
It looks very simple, indeed, and is rotationally invariant. 
All well-known difficulties of the front form
with rotational invariance seem to be absorbed 
by the factor $\sqrt{x(1-x)}$ in the mapping
Eq.(\ref{eq:psi}).
In \cite{Pau00b} it has been shown how to solve 
the equation numerically for spherical symmetry 
$\varphi(\vec p)=\varphi(p)$. 
Some further results are compiled in Table~\ref{eq:results}, below.
Here and below, masses and momenta are expressed 
in units of $u=350$~MeV, except when noted otherwise.
All parameter sets produce a lowest mass eigenvalue 
of $M^2=(139.57018\mbox{ MeV})^2$,
the mass (squared) of the $\pi^+$. Both the $M^2$ and the two first
exited states are stationary with respect to $\mu$.
The (unphysical) regularization parameter $\mu$ determines
itself from the solution (`renormalization'), 
for details see \cite{Pau00b}.

Next, before proceeding with the computation of the
form factor, a number of notational definitions 
are introduced, in terms of which the final
results turn out to be simple.
Once one has $\varphi(\vec p)$ 
in a parametrized form like Eq.(\ref{eq:ff}),
one can transform back to the variables $x$ and $\vec k _{\!\perp}$ 
and define
\begin{equation}
   Z(x,k_{\!\perp}) \equiv
   1 + \frac{p^2}{p_a^2} = 1+\frac{m^2}{p_a^2}
   \frac{(x-\frac{1}{2})^2} {x(1-x)} +
   \frac{\vec k_{\!\perp}^{\,2}} {4p_a^2x(1-x)} 
,\label{eq:Z}\end{equation}
as well as $X(x) \equiv Z(x,k_{\!\perp}=0)$.
The combination $1+p^2/m^2$ is trivially obtained from this 
by putting $p_a=m$.
The dimensionless parameters $s$ and $t$,
\begin{equation}
   s=\frac{m}{p_a}, \qquad\mathrm{and}\quad t^2=\frac{m^2-p_a^2}{m^2}
,\label{eq:s}\end{equation}
will govern the results below.
The form factor in Eq.(\ref{eq:4}) will be calculated
in the form
\begin{eqnarray}
   F_2(q_{\!\perp}^2) &=& 
   \int\limits_0^1 dx \int\limits_0^\infty dk_{\!\perp}^2\,
   \vert\psi(x,\vec k_{\!\perp})\vert^2 \,
   f(x,k_{\!\perp};q_{\!\perp})
,\label{eq:f}\\ \mathrm{with}\quad
   f(x,k_{\!\perp};q_{\!\perp}) &=&
   \int\limits_0^{2\pi} d\phi\ %
   \frac{e_1\psi(x,\vec k_{\!\perp}+(1-x)\vec q _{\!\perp})
       + e_2\psi(x,\vec k_{\!\perp}-  x  \vec q _{\!\perp})}
   {2\,\psi(x,\vec k_{\!\perp})}
.\end{eqnarray}
The function $f(x,k_{\!\perp};q_{\!\perp})$ contains all the 
difficulty in the problem, particularly
the integration over the angle $\phi$ 
between $\vec k_{\!\perp}$ and $\vec q_{\!\perp}$.
It depends only on the absolute value of the momentum transfer.
If one restricts to calculate only the derivative with respect 
to $q_{\!\perp}$, as in 
\begin{eqnarray}
   \langle r^2\rangle_2  &=& 
   \int\limits_0^1 dx \int\limits_0^\infty dk_{\!\perp}^2\,
   \vert\psi(x,\vec k_{\!\perp})\vert^2 \,g(x,k_{\!\perp})
,\label{eq:15}\\ \mathrm{with}\quad
   g(x,k_{\!\perp}) &=& -6 \left.\displaystyle
   \dot f(x,k_{\!\perp};q_{\!\perp})
   \right|_{q_{\!\perp}=0} =-6 \left.\displaystyle
   \frac{d}{d q_{\!\perp}^2} f(x,k_{\!\perp};q_{\!\perp})
   \right|_{q_{\!\perp}=0} 
,\label{eq:g}\end{eqnarray}
one deals with a function $g(x,k_{\!\perp})$ 
which is independent of $q_{\!\perp}$.

For calculating the probability amplitude according to Eq.(\ref{eq:Z2})
with a normalized wave function as in Eq.(\ref{eq:psi}), 
and the mean-square radius according to Eq.(\ref{eq:15}), 
one needs to evaluate three integrals, namely
\begin{eqnarray}
   \frac{4\pi}{3}p_a^2 N(s) \mathcal{N}^2 &=& 
   \int dx d^2k_{\!\perp} \psi^2(x,k_{\!\perp}) 
   ,\hskip7em(\ \equiv 1\ )
\nonumber\\
   2\pi p_a^2 P(s) \mathcal{N}\phantom{^2}&=&
   \int dx d^2k_{\!\perp} \psi\phantom{^2}(x,k_{\!\perp})
,\nonumber\\
   \pi S(s) \mathcal{N}^2 &=&
   \int dx d^2k_{\!\perp} \psi^2(x,k_{\!\perp}) g(x,k_{\!\perp})
   .\hskip3em(\ \equiv \langle r^2\rangle \ )
\label{eq:ACR}\end{eqnarray}
They are cast into three dimensionless functions
$N(s)$, $P(s)$, and $S(s)$, which depend only on the 
dimensionless parameter $s$ as defined in Eq.(\ref{eq:s}). 
Once they are known, the calculation of the probability amplitude
and the root-mean-square radius is easy. 
In particular, with
\begin{eqnarray}
   Z_2 &=&   \frac{f_\pi}{p_a}
   \frac{2\pi}{3} 
   \frac{\sqrt{N(s)}}{P(s)} =
   \frac{0.5665}{m}\ s\ \frac{\sqrt{N(s)}}{P(s)}
,\label{eq:Z2fin}\\
   R_2 &=& \frac{\sqrt{3}}{2p_a}\ \frac{\sqrt{S(s)}}{\sqrt{N(s)}}
,\hskip5em
   R_e  =  \frac{\sqrt{3}}{2p_a} 
,\label{eq:rmsfin}\end{eqnarray}
one can study them as dimensionless functions of $s$
for a fixed value of $m$.
\section{An approximate treatment}
\label{sec:3}
As will be seen below, the factor $\sqrt{1+p^2/m^2}$ 
in Eq.(\ref{eq:psi}) poses certain problems in the evaluation 
of the form factor and other integrals. 
In order to proceed pedagogically, 
this factor is first replaced here by unity,
by using 
\begin{equation} 
   \psi (x,k_{\!\perp}) \equiv
   \frac{\mathcal{N}}{\sqrt{x(1-x)}} \frac{1}{\left(1+ p^2/p_a^2\right)^\kappa}
,\label{eq:mod-rel}\end{equation}
with the two adjustable parameters $\kappa$ and $p_a$.
This is justified, of course, only 
if the Bohr momentum obeys $p_a^2 \ll m^2$, {\it i.e.} 
in the non-relativistic limit. But here $p_a/m \simeq 1.2$:
The quarks move highly relativistically! 

We disregard this objection, and proceed.
For to compute the mean-square radius, one needs to evaluate the
function $f$ as defined in Eq.(\ref{eq:f}).
For the parametrization of Eq.(\ref{eq:mod-rel}) one has thus
\begin{eqnarray*}
   f(x,k_{\!\perp};q_{\!\perp}) &=& 
   \frac{e_1}{2} Z^\kappa \int\limits_{0}^{2\pi}
   d\phi\ \left[1+\displaystyle 
   \frac{m^2}{p_a^2} \frac{(x-\frac{1}{2})^2}{x(1-x)}+
    \frac{\left(\vec k_{\!\perp}-(1-x)\vec q_{\!\perp}\right)^{\,2}}
    {4 p_a^2 x(1-x)}\right]^{-\kappa}
\nonumber\\ &+&
   \frac{e_2}{2} Z^\kappa \int\limits_{0}^{2\pi}
   d\phi\ \left[1+\displaystyle 
   \frac{m^2}{p_a^2} \frac{(x-\frac{1}{2})^2}{x(1-x)}+
    \frac{\left(\vec k_{\!\perp}-x\vec q_{\!\perp}\right)^{\,2}}
    {4 p_a^2 x(1-x)}\right]^{-\kappa}
.\end{eqnarray*}
It has two contributions, one from the quark ($\sim e_1$)
and one from the anti-quark ($\sim e_2$); they differ from each 
other by exchanging $x$ with $1-x$.
Since one is interested in the $\phi$-integration, 
one rewrites the equation as
\begin{equation}
   f(x,k_{\!\perp};q_{\!\perp}) = 
   Z^\kappa\int\limits_{0}^{\pi}d\phi\ \left[
   \frac {e_1}{\left( b_1 + c_1\cos\phi\right)^\kappa}+
   \frac {e_2}{\left( b_2 + c_2\cos\phi\right)^\kappa} \right]
,\label{eq:f-int}\end{equation}
where the coefficient functions $b$ and $c$ for the quark are given by
\begin{eqnarray}
  &&b_1 = Z+\frac{1-x}{x}\ \frac{q_{\!\perp}^{\,2}}{4p_a^2},\hskip3em
    c_1 = -\frac{1}{x}\frac{k_{\!\perp}}{p_a} \frac{q_{\!\perp}}{2p_a},
\label{eq:bc}\\
  &&\dot b_1 = \frac{1}{4\ p_a^2}    \ \frac{1-x}{x}, \hskip4em
   c_1\dot c_1 = \frac{1}{2\ p_a^2}\ \frac{1-x}{x}\ (Z-X)
.\label{eq:21}\end{eqnarray}
The functions $Z(x,k_{\!\perp})$ and $X(x)$ had been defined
in Eq.(\ref{eq:Z}). The coefficients for the anti-quark,
{\it i.e.} $b_2,c_2,\dot b_2$ and $\dot c_2$, 
are obtained from those for the quark
by exchanging $x$ with $1-x$.

However, one need not know explicitly $f(q_{\!\perp})$.
As a big advantage of the present aim to calculate $\langle r^2\rangle $, 
one needs the form factor only for very small $q_{\!\perp}$. 
Since $c_1$ and $c_2$ vanish for $q_{\!\perp}\rightarrow0$, 
one can expand Eq.(\ref{eq:f-int}) with $c$ 
\underline{before integration}. 
The contribution from the quark becomes then
\begin{equation}
   f_1(x,k_{\!\perp};q_{\!\perp}) = e_1 \frac{Z^\kappa}{b_1^\kappa}  
   \int\limits_{0}^{\pi}d\phi\ \left[1 - 
   \kappa \frac{c_1}{b_1} \cos\phi +
   \frac{\kappa(\kappa+1)}{2}\frac{c_1^2}{b_1^2} \cos^2\phi + \dots
   \right]
.\label{eq:expansion}\end{equation}
Note that the term of first order in ${c_1}/{b_1}$ 
vanishes upon integration, and that the term of 
second order is integrated trivially. 
Taking the derivative with respect to $q_{\!\perp}^2$
according to Eq.(\ref{eq:g}), one is left with 
\[
   g_1(x,k_{\!\perp}) = 6 e_1 \pi \frac{Z^\kappa}{b_1^\kappa}  
   \left[\kappa\frac{\dot b_1}{b_1} -
   \frac{\kappa(\kappa+1)}{2}\frac{c_1 \dot c_1}{b_1^2}
   \right]_{b_1=Z}
.\]
After inserting the derivatives according to Eq.(\ref{eq:21}) one gets
\begin{equation}
   g_1(x,k_{\!\perp}) = \frac{3 \pi\kappa }{2p_a^2}  
   \left[-\frac{\kappa}{Z} +(\kappa+1) \frac{X}{Z^2}\right]
   \ e_1 \frac{1-x}{x}
.\label{eq:g4}\end{equation}
The total $g=g_1+g_2$ is thus proportional to
\[
   e_1\frac{1-x}{x} + e_2\frac{x}{1-x} = 
   \frac{\left[(e_1+e_2)x^2 + e_1(1-2x)\right]}{x(1-x)} 
   \longmapsto \frac{x^2}{x(1-x)}     
.\]
The great simplification occurs since $e_1+e_2=1$, and 
since both $X$ and $Z$ are symmetric under the exchange
of $x$ and $1-x$, such that $1-2x$ vanishes upon integration.
Finally, the $g$-function of Eq.(\ref{eq:g}) 
for the semi-relativistic wave function, Eq.(\ref{eq:mod-rel}),
becomes for arbitrary $\kappa$
\begin{equation}
   g(x,k_{\!\perp}) = \frac{3\pi\kappa}{2p_a^2}  
   \left[-\frac{\kappa}{Z} +(\kappa+1) \frac{X}{Z^2}\right]
   \ \frac{x^2}{x(1-x)}
.\label{eq:gfin}\end{equation}
One should emphasize that $g(x,k_{\!\perp})$ was obtained 
without explicitly evaluating Eq.(\ref{eq:f-int}) and 
that the results for $\kappa=2$ and $\kappa=3/2$ agree  
with the expressions in \cite{PaM01}, 
where the limit $q_{\!\perp}\rightarrow0$ had been 
taken \underline{after integration}.

Having $g$ from Eq.(\ref{eq:gfin}) 
and the wave function from Eq.(\ref{eq:mod-rel}),
one can calculate the three integrals as defined in 
Eq.(\ref{eq:ACR}), particularly
\begin{eqnarray}
   N(s) &=& \frac{3}{4p_a^2} \frac{1}{\mathcal{N}^2} 
   \int dx\,dk_{\!\perp}^2\ \psi^2(x,k_{\!\perp})
,\hskip1em
   P(s) = \frac{1}{2p_a^2} \frac{1}{\mathcal{N}\phantom{^2}}
   \int dx\,dk_{\!\perp}^2\ \psi(x,k_{\!\perp})
,\nonumber\\
   S(s) &=& \frac{1}{\pi} \frac{1}{\mathcal{N}^2}
   \int dx\,dk_{\!\perp}^2\ \psi^2(x,k_{\!\perp})\ g(x,k_{\!\perp})
.\label{eq:defACR}\end{eqnarray}
Before evaluating them, it is advantageous
to change the integration variables $x$ and $k_{\!\perp}^2$
to $z$ and $u$, respectively, with
\begin{equation}
   z=2x-1, \qquad \mbox{and}\quad u=\frac{k_{\!\perp}^2}{m^2}
.\label{eq:z}\end{equation}
$Z(x,k_{\!\perp})$ and $X(x)$ as defined in Eq.(\ref{eq:Z})
are then rewritten as
\begin{equation}
   Z(z,u) = s^2\frac{\left[1+(z^2-1)t^2+u\right]}{\left[1-z^2\right]}
   , \quad 
   X(z) = s^2\frac{\left[1+(z^2-1)t^2\right]}{\left[1-z^2\right]}
.\label{eq:Zz}\end{equation}
Evaluating Eq.(\ref{eq:defACR}) for $\kappa=2$ 
gives the normalization function 
\begin{eqnarray}
   N(s) &=& \frac{3}{s^6}
   \int\limits_{-1}^{+1} \frac{dz}{2}\ \int\limits_0^\infty du\ %
   \frac{\left[1-z^2\right]^3}{\left[1+(z^2-1)t^2+u\right]^4} 
\label{eq:25}\\ &=& 
   \frac{-8-10s^2+3s^4 + 3b(s)s^2 (8- 4s^2+ s^4)}{8(s^2-1)^3} 
.\end{eqnarray}
Here and below the abbreviation
$b(s)= \mathrm{arctan}( \sqrt{s^2-1})/\sqrt{s^2-1}$  
is used.
The major labor of the present work is hidden in the evaluation
of integrals such as Eq.(\ref{eq:25}) and to express them as 
explicit functions of $s$.
In order not to load the paper with straightforward formalism, 
the arithmetics are suppressed here, as a rule, the more as they
can be produced also by Mathematica.
Care was taken that no typos do occur.~--- 
For the probability function one obtains
\begin{equation}
   P(s) = \frac{1}{s^2}
   \int\limits_{-1}^{+1} \frac{dz}{2}\ \int\limits_0^\infty du\ %
   \frac{\left[1-z^2\right]^{\frac{3}{2}}}{\left[1+(z^2-1)t^2+u\right]^2} 
   = \frac{\pi}{4}
   \frac{1 - 3s^2 + 2 s^3}{(s^2-1)^2}
,\label{eq:nps4}\end{equation}
and the size function finally is
\begin{eqnarray}
   &&S(s) = \frac{12}{s^8}
   \int\limits_{-1}^{+1} \frac{dz}{2}\ \int\limits_0^\infty du\ %
   \left[1+z\right]^2 \left[1-z^2\right]^3
   \frac{\left[1+(z^2-1)t^2-2u\right]}{\left[1+(z^2-1)t^2+u\right]^6} 
\label{eq:S4}\\ &&= 
   \frac{-136 +72s^2 -56s^4 +15s^6 
   +3b(s)( 32 -16s^2 +36s^4 -22s^6 +5s^8)}{40(s^2-1)^4}
.\nonumber\end{eqnarray}
These functions are plotted below 
as function of $s$ in the region of interest $s\sim 1$. 
For $\kappa=\frac{3}{2}$ they are given in \cite{PaM01}.
\section{The full treatment}
\label{sec:4}
Having been so explicit in the previous section,
we can be more compact when restoring the factor 
$\sqrt{1+{\vec p ^{\;2}}/{m^2}}$
in the wave function 
\begin{equation}
   \psi(x,k_{\!\perp}) \equiv
   \frac{\mathcal{N}}{\sqrt{x(1-x)}} 
   \frac{\sqrt{1+p ^2/m^2}}{\left(1+ p^2/p_a^2\right)^2} \equiv
   \frac{\mathcal{N}}{\sqrt{x(1-x)}} 
   \frac{Y ^{\frac{1}{2}}(x,k_{\!\perp})}{Z^2(x,k_{\!\perp})}  
.\label{eq:ful-rel}\end{equation}
The origin of the coefficient function $Y(x,k_{\!\perp})$ is obvious, 
{\it i.e.} 
\begin{equation}
   Y(x,k_{\!\perp}) \equiv 1 + \frac{p^2}{m^2} = 
   \frac{m^2 + \vec k_{\!\perp}^{\,2}} {4m^2x(1-x)} 
,\qquad
   Y(z,u) =  \frac{[1+u]}{[1-z^2]}
.\label{eq:ZT}\end{equation}
Correspondingly one has $U(x) \equiv Y(x,k_{\!\perp}=0)$
and $U(z) \equiv Y(z,u=0)$. 
The coefficient function $Z$ was defined in Eq.(\ref{eq:Z}).
Note that the full relativistic wave function interpolates 
to some extent between $\kappa=2$ and $\kappa=\frac{3}{2}$ 
in the previous section: Substituting  $m$ by $p_a$
gives the same wave function as for $\kappa=\frac{3}{2}$,
while putting $m\rightarrow \infty$ gives the wave function for $\kappa=2$.

For to compute the form factor,
one needs to evaluate $f$ as defined in Eq.(\ref{eq:f}).
The analogue of Eq.(\ref{eq:f-int}) is
\begin{equation}
   f(x,k_{\!\perp};q_{\!\perp}) = \frac{Z^2}{Y^{\frac{1}{2}}}
   \int\limits_{0}^{\pi}d\phi\ \left[
   e_1\frac {
   \left(B_1 + C_1\cos\phi\right)^{\frac{1}{2}}}
      {\left( b_1 + c_1\cos\phi\right)^2}+
   e_2\frac {
   \left(B_2 + C_2\cos\phi\right)^{\frac{1}{2}}}
      {\left( b_2 + c_2\cos\phi\right)^2} \right]
.\label{eq:ftilde}\end{equation}
The coefficient functions $b_i$ and $c_i$ had been given 
above, and
\begin{eqnarray}
  &&B_1 = 
  Y+\frac{1-x}{x}\ \frac{q_{\!\perp}^{\,2}}{4m^2},\hskip3em
    C_1 = -
    \frac{1}{x}\frac{k_{\!\perp}}{m} \frac{q_{\!\perp}}{2m},
\label{eq:bctilde}\\
  &&\dot{B_1} = \frac{1}{4\ m^2}    
  \ \frac{1-x}{x}, \hskip4em
  C_1\dot{C_1} = 
  \frac{1}{2\ m^2}\ \frac{1-x}{x}\ (Y-U)
.\label{eq:bcdottilde}\end{eqnarray}
Again, the anti-quark coefficients are obtained 
by exchanging $x$ with $1-x$.

The analogue of Eq.(\ref{eq:expansion})
is now much more complicated 
but still straightforward.
In full analogy with Eq.(\ref{eq:gfin})
one gets finally
\begin{equation}
   g(x,k_{\!\perp}) = \frac{3\pi}{p_a^2}  
   \left[-\frac{2}{Z} + \frac{3X}{Z^2} + \frac{1}{s^2}
   \left(\frac{7}{8Y} - \frac{X}{ZY} -
   \frac{U}{8Y^2} \right) \right]
   \ \frac{x^2}{x(1-x)}
.\label{eq:gfingen}\end{equation}
Note that this equation is consistent with the previous results:
Omitting the terms in the round bracket (thus $m\rightarrow \infty$)
agrees with Eq.(\ref{eq:gfin}) for $\kappa=2$,
while formally putting $s=1$ and $Y=Z$
reproduces it for $\kappa=\frac{3}{2}$. 

Having $g$ and the wave function,
one can evaluate the three integrals as defined in 
Eq.(\ref{eq:ACR}).
The normalization function becomes
\begin{eqnarray}
   N(s) &=& \frac{3}{s^6}
   \int\limits_{-1}^{+1} \frac{dz}{2}\int\limits_0^\infty du\ %
   \frac{\left[1-z^2\right]^2
   \ \left[1+u\right]}{\left[1+(z^2-1)t^2+u\right]^4} 
\nonumber\\ &=& 
   \frac{ 4-4s^2+3s^4+3b(s)s^4(-2+s^2)}{8s^2(s^2-1)^2} 
,\label{eq:Nrel}\end{eqnarray}
and the probability function is
\begin{equation}
   P(s) =
   \int\limits_{-1}^{+1} \frac{dz}{2}\ \int\limits_0^\infty %
   \frac{du\ \left[1-z^2\right]\ \left[1+u\right]^{\frac{1}{2}}}
   {\left[1+(z^2-1)t^2+u\right]^2} 
   \simeq 
   \frac{s^2 + b(s)(s+s^2-s^3-2s^4)}{2(1-s^2)}
.\label{eq:Prel}\end{equation}
Its integration is non-trivial, see App.~\ref{app:1}.
The size function $S(s)$ is now much more complicated
than in the previous section, but after elementary
manipulations one gets 
\begin{eqnarray}
   S(s) &=& \frac{12}{s^8}
   \int\limits_{-1}^{+1} \frac{dz}{2}\ \int\limits_0^\infty du\ %
   \left[1+z\right]^2 \left[1-z^2\right]^2 \left(
   -\frac{2\left[1+u\right]}{\left[1+(z^2-1)t^2+u\right]^5}\right.
\nonumber\\ &&+\left.
    \frac{3\left[1+u\right]\left[1+(z^2-1)t^2\right]}
    {\left[1+(z^2-1)t^2+u\right]^6}
   +\frac{7}{8\left[1+(z^2-1)t^2+u\right]^4}\right.
\nonumber\\ &&-\left.
    \frac{\left[1+(z^2-1)t^2\right]}{\left[1+(z^2-1)t^2+u\right]^5}
   -\frac{1}{8\left[1+u\right]\ \left[1+(z^2-1)t^2+u\right]^4}
   \right) 
\label{eq:Srel}\\ &=& 
   \frac{ 34 +37s^2 -41s^4 +15s^6 + 
   3b(s)(- 8 -16s^2 +21s^4 -17s^6+5s^8)}{40s^2(s^2-1)^3}
.\nonumber\end{eqnarray}
This completes part of the aim of the present work, namely to calculate explicitly 
the mean-square radius $\langle r^2\rangle$
and the probability amplitude $Z_2$,
see Eqs.(\ref{eq:Z2fin}) and (\ref{eq:rmsfin}).
\section{Discussion}
\label{sec:5}

In a non-relativistic system one calculates the mean-square
radius comparatively cheaply by Fourier transforming
a momentum space wave function to configuration space,
and taking the expectation value of $r^2$.
Taking for simplicity the function $\varphi(p)$ in Eq.(\ref{eq:ff})
as such a `wave function' with all due precaution,
precisely this was intended in \cite{Pau00b}, 
with the resulting non-relativistic estimate 
$R_e =  {\sqrt{3}}/{2p_a}$ given in Eq.(\ref{eq:rmsfin}).
It hurts to admit, that a missing factor 2 in \cite{Pau00b} 
was unraveled only in the course of the present work, 
particularly in \cite{PaM01}.
But now, we are in a better shape: Not only are we able to
calculate the mean-square radius in the almost same way
as an experimentalist performs the measurement,
but we can calculate it analytically, 
irrespective of whether one deals with a non-relativistic system or not.
How does the discrepancy  behave?

\begin{figure} [t]
\begin{minipage}[t]{67mm}
  \resizebox{0.99\textwidth}{!}{\includegraphics{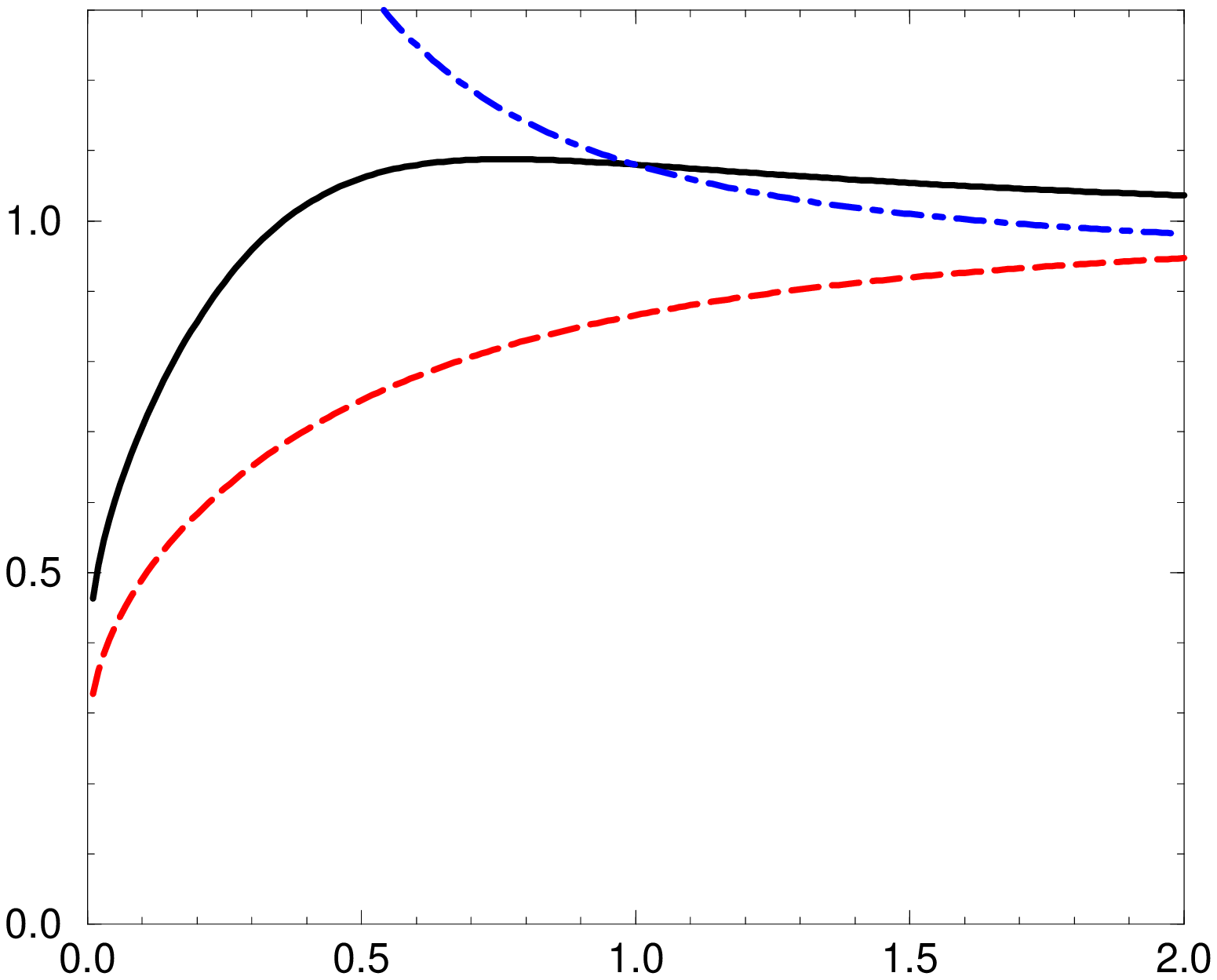}}
  \caption{\label{fig:nos243} 
    The discrepancy $d = R_e/R_2$ 
    is plotted versus $s$ by the solid line. 
    It is compared to the semi-relativistic case by the dashed 
    ($\kappa=2$) and the dashed-dotted line ($\kappa=3/2$).
}\end{minipage} 
\ \hfill
\begin{minipage}[t]{67mm}
  \resizebox{0.99\textwidth}{!}{\includegraphics{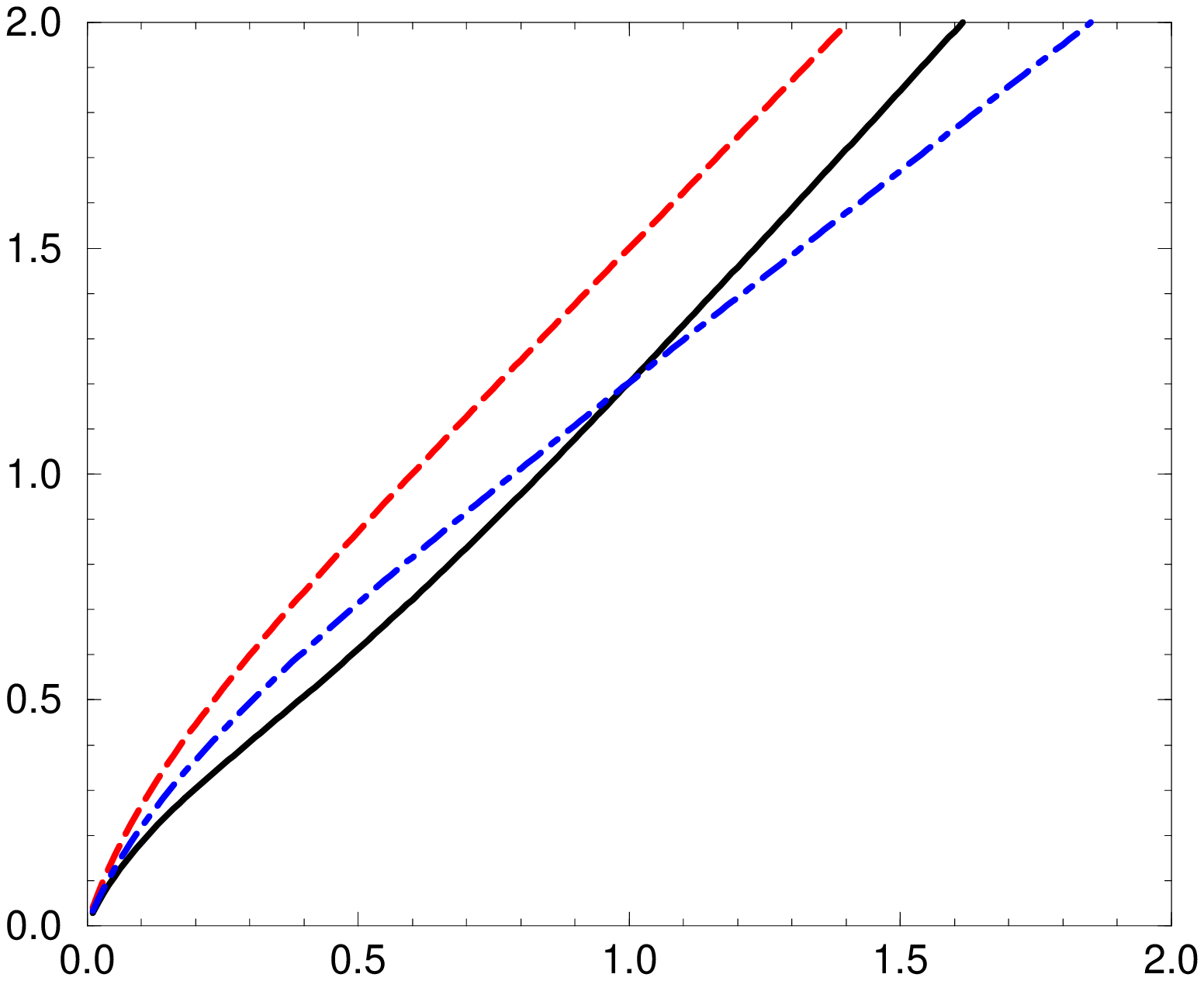}}
  \caption{\label{fig:rms} 
    The rms-radius of the pion in units of 0.67 fm
    is plotted versus $s$ by the solid line.
    It is compared with the approximate cases by the dashed ($\kappa=2$) 
    and the dashed-dotted line ($\kappa=3/2$).
}\end{minipage} 
\end{figure}

The discrepancy $d = R_e/R_2 = \sqrt{{N(s)}/{S(s)}}$ 
is plotted versus $s=m/p_a$ in Fig.~\ref{fig:nos243}. 
One observes that the curve for
the approximate wave function of Eq.(\ref{eq:mod-rel})
for $\kappa=2$ 
approaches the limiting value $d=1$ rather
quickly from below for growing $s$.
Note that there is a long way to go until $s\sim 137$,
the value for a Bohr atom with Bohr momentum
$p_a\sim m\alpha$. 
This can also be seen from the analytic formulas,
Eqs.(\ref{eq:25}) and (\ref{eq:S4}).
The case for $\kappa=3/2$ has also a limit, 
however uninteresting in this context.
Fig.~\ref{fig:nos243} displays as well that the
correct wave function from Eq.(\ref{eq:ful-rel})
approaches the limit from above.
The discrepancy exceeds rarely some few percent
for values of $s$ down to $s \sim 0.3$.
That the non-relativistic estimate is so accurate
was a surprise. 
Ultimately, for $p_a\rightarrow\infty$
thus $s\rightarrow 0$, for a point like system,
the discrepancy and thus the size drops to zero 
even faster than the non-relativistic estimate.
This can be seen also in Fig.~\ref{fig:rms},
where the size is plotted for the various cases 
considered in this work.

%
%
\begin{table}
\caption{\label{eq:results}
   Compilation of results.
}
\begin{tabular}{lcc@{\hspace{3em}}ccccc@{\hspace{2.5em}}cc} 
\hline\hline
    set &$\alpha$& $m$&$p_a$& $p_a/m$ & $s$ & $d$ &$Z_2$& $R_e$ & $R_2$\\ 
    { } &   { }  & MeV&MeV/c&   { }   & { } & { } & { } &  fm  &  fm  \\
\hline
 $\# 1$ & 0.690  & 406& 515 & 1.268   &0.789&0.919&0.392& 0.33 & 0.30 \\
 $\# 2$ & 0.682  & 301& 378 & 1.255   &0.797&0.919&0.524& 0.45 & 0.42 \\
 $\# 3$ & 0.672  & 196& 234 & 1.192   &0.839&0.920&0.774& 0.73 & 0.67 \\
\hline\hline
\end{tabular}
\end{table}

Table~\ref{eq:results} summarizes the results for solving 
Eq.(\ref{eq:bsEq}) numerically, in the manner described in \cite{Pau00b}.
Set $\# 1$ quotes the results of \cite{Pau00b}.
The entry $R_e=0.33\mbox{ fm}$ for the non-relativistic
estimate reflects the factor 2 mentioned.
The calculation of the exact root-mean-square radius $R_2=0.30\mbox{ fm}$
was one of the motivations to get this work started.
The discrepancy $d=0.92$ is amazingly small.
Two additional calculations have been performed,
one with the constraint $R_2=\sqrt{\langle r^2\rangle_{exp}}=0.67\mbox{ fm}$
in order to pin down the quark mass $m$, and the other
with $R_2=0.42$~fm, referred to as set $\# 3$ and $\# 2$, respectively.
The quark mass $m$ determined from the constraint on the rms
is almost linear to $1/R_2$. But the width of the reduced wave
function changes almost in proportion, such that $p_a/m\sim 1.2$
(or $s\sim0.8$) and thus $d$ is almost independent of the case. 
No explanation for this numerical observation can be offered.

Table~\ref{eq:results} includes also the probability amplitude $Z_2$, 
as calculated with Eq.(\ref{eq:Z2fin}).
Early estimates of this quantity from  Brodsky, Huang and Lepage 
\cite{BHL80,BHL81} imply that the pion is roughly 1/4 of the time 
in the $q\bar q$-Fock state, 
but it is the first time, 
that such a number is calculated from a wave function.
The numbers in the table imply 1/2 to 1/5 for this quantity,
depending on the case, in rough agreement with \cite{BHL80,BHL81}.

As a matter of fact, Brodsky, Huang and Lepage emphasize 
also that 2-particle Fock-state is {\em smaller} or more compact
than the pion according to Eq.(\ref{eq:1}).
The estimate they give in \cite{BHL81} (on pg. 20) is
{\it mutatis mutandis} $R_2\simeq \widetilde Z_2\cdot(0.7\mbox{ fm})$,
where $\widetilde Z_2^2$ 
(not to be confused with the $Z_2$ of Eq.(\ref{eq:Z2fin}))
is the probability to find a bare quark in a dressed quark.
It is certainly less than 1, 
but neither measured nor computed, thus far.
One concludes that a fit to the electro-magnetic size
as in set $\# 3$ is the least favorable of the three sets 
in Table~\ref{eq:results}.
Set $\# 1$ corresponds to a `hard core radius' 
of about $R_2\sim0.3\mbox{ fm}$, corresponding to a 
$\widetilde Z_2\sim\frac{1}{2}$.
Some people favor a hard core radius of 0.4~fm, which was the
motivation for calculating with set $\# 2$.

\section{The distribution function}
\label{sec:6}
In a recent experiment, Ashery \cite{Ash00,AshColl00} has provided the first 
direct measurement of the pion light-cone wave function (squared):
A high energy pion dissociates diffractivly on a heavy nuclear target. 
In a coherent process the quark and the antiquark break apart 
and hadronize into two jets.
Their momentum distribution carries information on
the quarks momentum distribution in the pion. 
By and large, Ashery's results agree with 
the $x(1-x)$-distribution predicted long ago 
by Lepage and Brodsky \cite{leb80}
and by Efremov and Radyushkin \cite{EfR80}, 
by solving a perturbative QCD evolution equation in the limit
of large momentum transfer $Q^2\rightarrow\infty$.
Ashery has fitted his data by
a linear superposition with the $x(1-x)(2x-1)^2$-distribution 
of Chernyak and Zhitnitsky \cite{ChZ84},
which is believed to be reliable in the opposite limit 
$Q^2\rightarrow0$. 

Although Ashery's experiment cannot be analyzed in terms of 
the distribution function alone, it is interesting to ask what
the present work predicts for it.~--
By convenience, the definition in Eq.(\ref{eq:Phi}) is split up 
into a normalization factor $N_{\Phi}$
and the reduced distribution function 
$\phi(x,k_t^2)$ , {\it i.e.}
$\Phi_{q\bar q}(x,k_t^2) = N_{\Phi} \phi(x,k_t^2)$, with
\begin{eqnarray}
   N_{\Phi} &=& 4\pi \mathcal{N} Z_2 \frac{p_a^4}{m^2} =
   f_{\pi} \frac{4\pi}{P(s)} \frac{p_a^2}{m^2} \sqrt{\frac{\pi}{n_c}}
,\label{eq:phinorm}\\
   \phi(x,k_t^2) &=& 2m\,x(1-x)\int_0^{k_t^2}\!dk_{\!\perp}^2
   \ \frac{\left[k_{\!\perp}^2+m^2\right]^{\frac{1}{2}}}
   {\left[k_{\!\perp}^2+a^2\right]^2}
,\label{eq:distDef}\\  
  a^2 &=& m^2 + 4x(1-x)\left(p_a^2-m^2\right)
.\label{eq:a2Def}\end{eqnarray}
Note that $a^2=a^2(x)$ actually is a function of $x$ and that 
$f_{\pi}$ is the carrier of the dimension.
The factor of $2m$ is inserted into the definition of $\phi$
to have a dimensionless function.
In Eq.(\ref{eq:phinorm}), $\mathcal{N}$ and $Z_2$
have been substituted according to 
Eqs.(\ref{eq:ACR}) and (\ref{eq:Z2fin}), respectively.
Brodsky has emphasized repeatedly \cite{BroPauPin98}
that one could normalize the wave function also by the weak 
pion decay amplitude, Eq.(\ref{eq:Z2}). 
It is gratifying to see that the normalization function $N(s)$ 
cancels in Eq.(\ref{eq:phinorm}) even without taking special 
care of that.
Evaluating the integral in Eq.(\ref{eq:distDef}), 
one gets for $p_a\ge m$ 
\begin{eqnarray}
  \phi(x,k_t^2) &=& x(1-x)\left[
  \frac{2m^2}{a^2} - 2m
  \frac{\left[k_t^2+m^2\right]^\frac{1}{2}}{\left[k_t^2+a^2\right]^2}\right.
\nonumber\\ 
  &&+ \left.\frac{2m}{\sqrt{a^2-m^2}}
  \left(\mathrm{arctg}\sqrt{\frac{a^2+k_t^2}{a^2-m^2}}-
  \mathrm{arctg}\sqrt{\frac{a^2}{a^2-m^2}}\right) 
  \right]
,\label{eq:relDist}\end{eqnarray}
and for $p_a\le m$ 
\begin{eqnarray}
  \phi(x,k_t^2) &=& x(1-x)\left[
  \frac{2m^2}{a^2} + \frac{m}{\sqrt{m^2-a^2}}
  \ln{\left(\frac{\sqrt{k_t^2+m^2}-\sqrt{m^2-a^2}}
  {\sqrt{k_t^2+m^2}+\sqrt{m^2-a^2}} \right)} \right.
\nonumber\\ && \hskip7em +
  \left.\frac{m}{\sqrt{m^2-a^2}}
  \ln{\left( \frac {m+\sqrt{m^2-a^2}}{m-\sqrt{m^2-a^2}} \right)}
  \right]
.\end{eqnarray}
The quadratures are elementary: One checks them by differentiating
the latter two equations with respect to $k_t^2$.

%
\begin{figure}[t]
\begin{minipage}[t]{67mm}
  \resizebox{0.99\textwidth}{!}{\includegraphics{{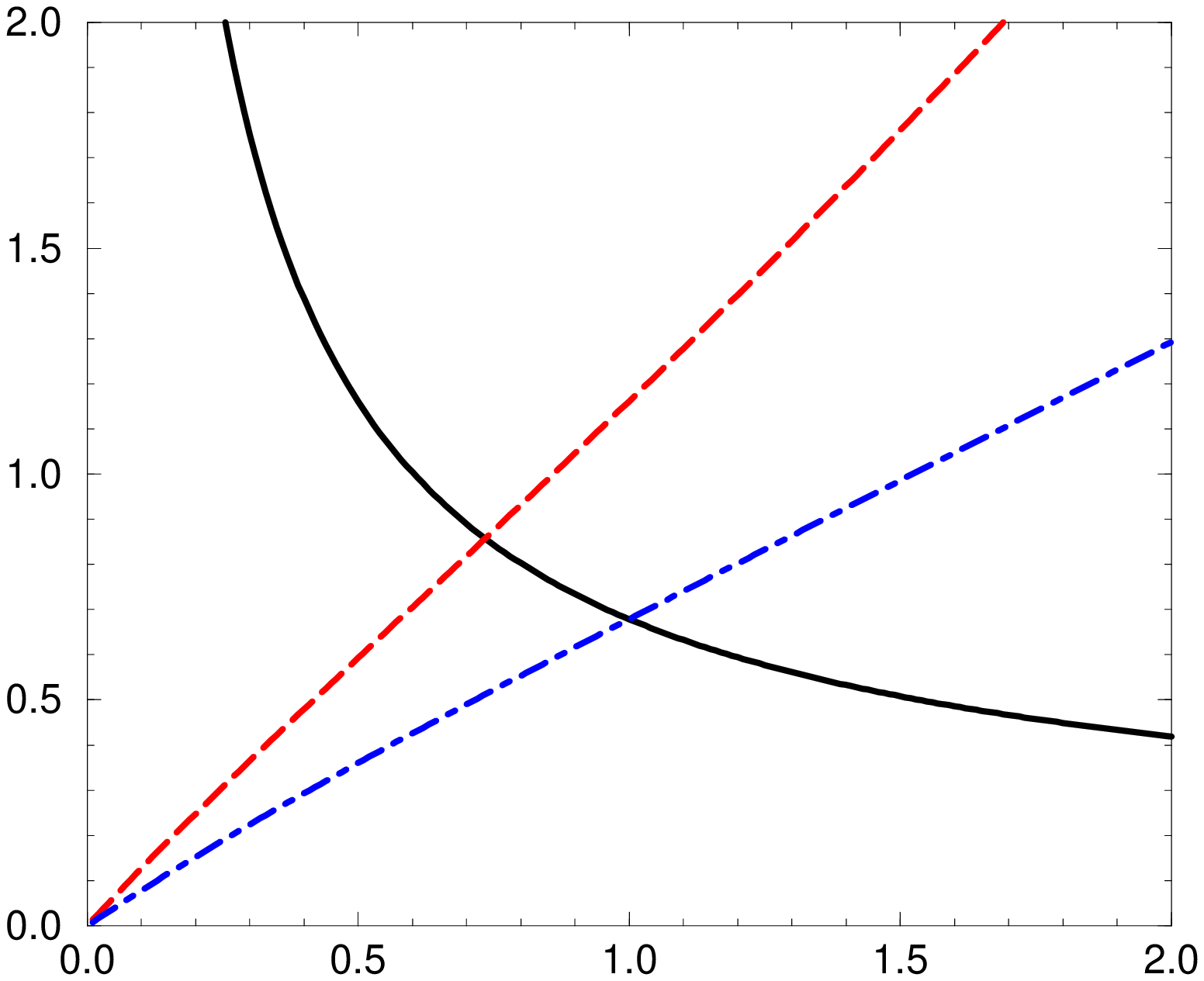}}}
  \caption{\label{fig:Z2} 
    The Fock-state probability amplitude $Z_2$ is plotted
    versus $s$ by the solid line. 
    It is compared to the approximate cases by the dashed 
    ($\kappa=2$) and the dashed-dotted line ($\kappa=3/2$).
}\end{minipage}
\ \hfill
\begin{minipage}[t]{67mm}
  \resizebox{0.99\textwidth}{!}{\includegraphics{{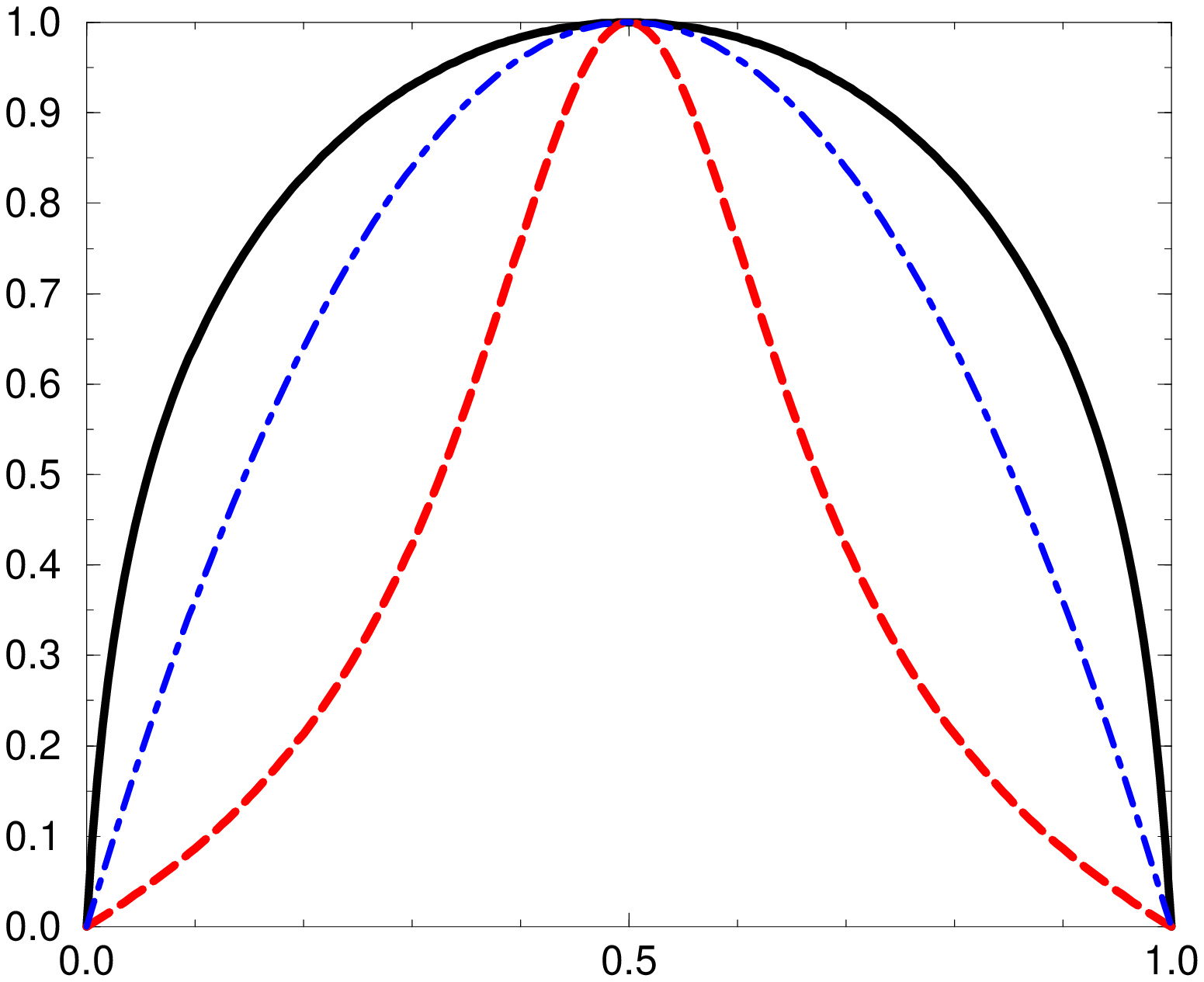}}}
  \caption{\label{fig:dist} 
    The reduced asymptotic distribution function $\phi_{as}(x)$
    as given by Eq.(\protect{\ref{eq:asyphi}}).
    The full line refers to $p_a=3m$,
    the dash-dotted line to $p_a=m$, and
    the dashed line to $p_a=m/3$.
}\end{minipage} 
\end{figure}
%
In the limit $k_t\to\infty$, one gets from this the 
asymptotic distribution function 
\begin{eqnarray}
  \phi_\mathrm{as}(x) &=& x(1-x)\cdot
  \cases{\displaystyle \frac{2m^2}{a^2}+ \frac{2m}{\sqrt{a^2-m^2}} 
         \mathrm{arctg}\sqrt{\frac{a^2-m^2}{a^2}},
            &for $p_a>m$,     \cr
         \displaystyle 4, &for $p_a=m$,     \cr
         \displaystyle \frac{2m^2}{a^2} + \frac{2m}{\sqrt{m^2-a^2}}
         \mathrm{Arth}\sqrt{\frac{m^2-a^2}{m^2}} , 
	    &for $p_a<m$.\cr }
\label{eq:asyphi}\end{eqnarray}
One notes first that the Lepage-Brodsky-Efremov-Raduyshkin limit
(LeBER) is reproduced by the overall factor $x(1-x)$.
For $p_a=m$, we reproduce it even exactly.
One should however not be surprised by deviations from this limit,
since hadronic scales like $m$ or $p_a$ had been omitted
consistently in the derivation \cite{leb80,EfR80}, 
as producing terms of order unity.
On the other hand, such terms can generate significant 
deviations and sometimes can change even the qualitative behaviour,
as will be discussed next.

In the familiar bound-state problems, the constituents
have a momentum whose mean ($\sim p_a$) is significantly
smaller than their mass.
As shown in Fig.~\ref{fig:dist},
such non-relativistic systems with $p_a^2\ll m^2$
have an asymptotic distribution which is (much) narrower
than the LeBER distribution $x(1-x)$.
In the extreme {\em non}-relativistic limit 
(as for example in a $\mu\bar e$-atom with equal masses), 
the distribution function will be peaked very sharply  
at $x=\frac{1}{2}$. 
The distribution functions will be plotted here 
in units of the peak value.
They are functions only of $s=m/p_a$, 
and not of $m$ and $p_a$ separately.
The reason is that the dimension of $\Phi$ is carried
by the pion decay constant, as mentioned.

We are less familiar with bound-state systems in which
the constituents have a mean-momentum larger than the mass. 
Such systems are hypothetical. Let us refer to them as
relativistic bound-state systems, characterized by $p_a^2> m^2$. 
As displayed in Fig.~\ref{fig:dist},
a (highly) relativistic system has a distribution function
(much) broader than the LeBER limit.
In particular, it is flatter around $x=\frac{1}{2}$.
In the limit $p_a\to\infty$, it becomes completely
flat, which reminds one to the point-like Schwinger/t'Hooft bosons
in 1-space and 1-time dimension \cite{BroPauPin98}.
%
\begin{figure}[t]
\begin{minipage}[t]{67mm}
  \resizebox{0.99\textwidth}{!}{\includegraphics{{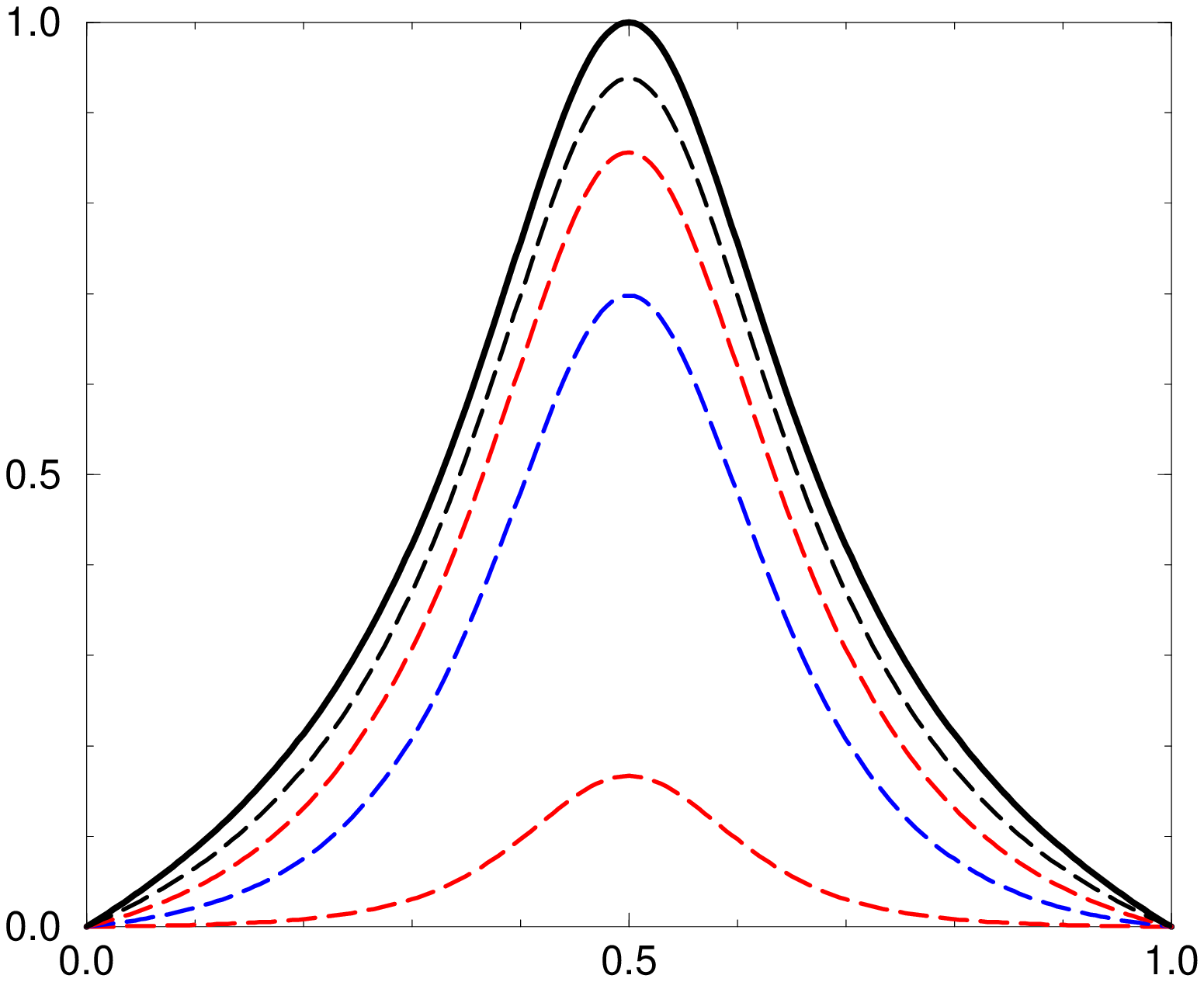}}}
  \caption{\label{fig:phiknon}
    The distribution function $\phi(x,k_t^2)$ 
    with $p_a=m/3$ is plotted versus $x$ for 
    $k_t=$ 32.6, 131, 262, 588 MeV by the four dashed lines, respectively.
    The solid line gives the asymptotic distribution function $\phi_{as}(x)$. 
    All functions are plotted in units of the peak value
    $\phi_{as}(\frac{1}{2})=$ 5.435. 
}\end{minipage}
\ \hfill
\begin{minipage}[t]{67mm}
  \resizebox{0.99\textwidth}{!}{\includegraphics{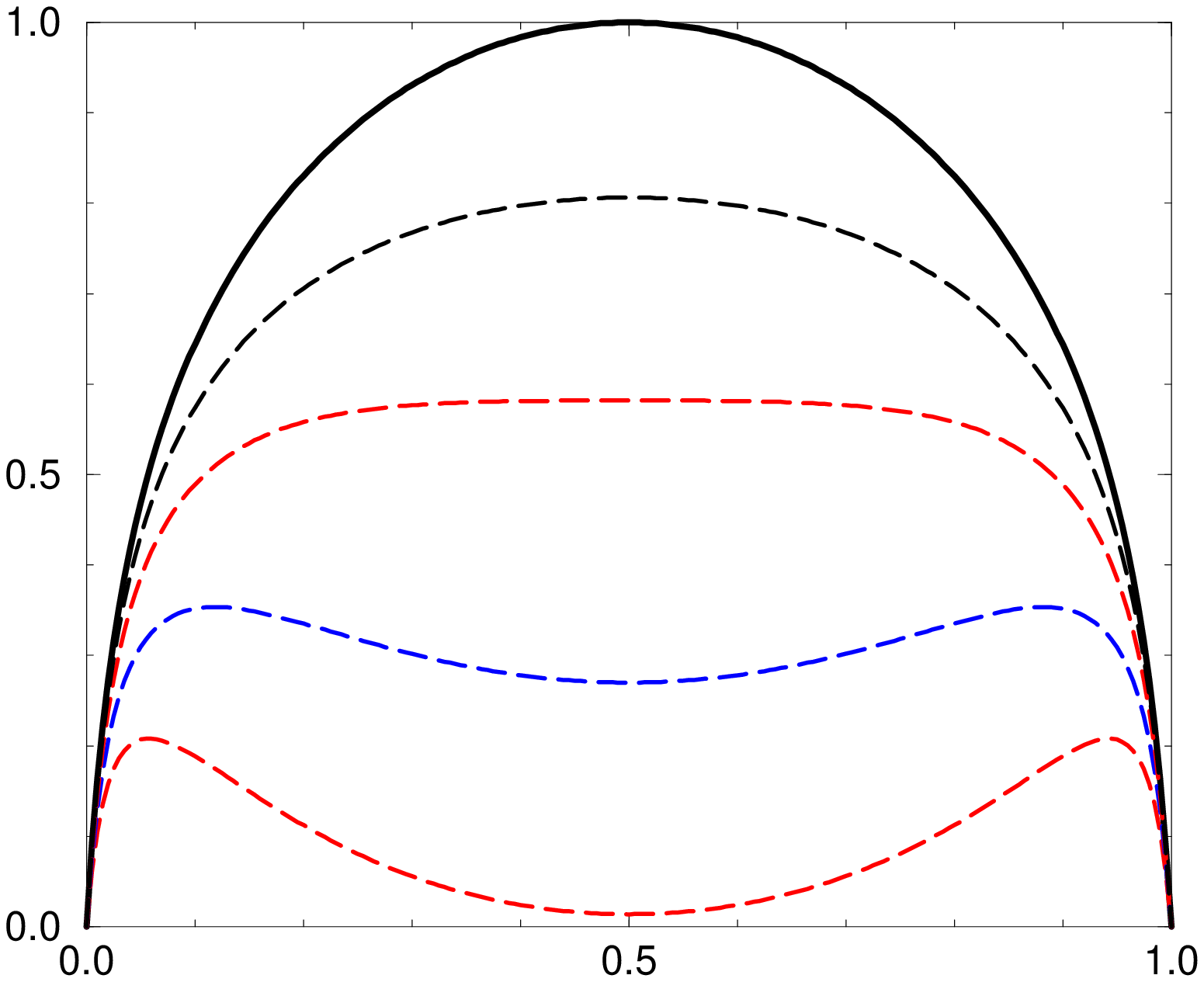}}
  \caption{\label{fig:phikrel} 
    The distribution function $\phi(x,k_t^2)$ 
    with $p_a=3m$ is plotted versus $x$ for 
    $k_t=$ 0.647, 1.18, 2.35, 5.29 GeV by the four dashed lines, respectively.
    The solid line gives the asymptotic distribution function $\phi_{as}(x)$. 
    All functions are plotted in units of 
    $\phi_{as}(\frac{1}{2})=$ 0.1892. 
}\end{minipage} 
\end{figure}

Similar considerations hold true also for finite values of $k_t$.
In Figs.~\ref{fig:phiknon} and \ref{fig:phikrel} 
the distribution function $\phi(x,k_t^2)$ is plotted 
for a non-relativistic
and a relativistic bound-state system, respectively,
for different values of $k_t$.
One should emphasize that the relativistic case develops `ears'
for sufficiently small $k_t$, which remind to the 
Chernyak-Zhitnitsky distribution.
These are absent in the non-relativistic case. 

\section{Summary and conclusion}
\label{sec:7}
The essential progress of this work is the consequent application 
of the light-cone wave function $\psi(x,\vec k_{\!\perp})$ 
in the parametrization of Eq.(\ref{eq:psi}), 
{\it i.e.} in terms of a reduced wave 
function $\varphi(x,\vec k_{\!\perp})$. 
The reduced wave function in turn is parametrized in Eq.(\ref{eq:ff})
like a Coulomb wave function in momentum space. 
The only adjustable parameter is
the analogue of the Bohr momentum $p_a$.
It is adjusted to a numerical solution of a mock-up,
the $\uparrow\downarrow$-model.
Both steps combined allow to calculate many observables, particularly  
the mean-square radius $\langle r^2 \rangle_2$, 
the  probability amplitude $Z_2$, and
the distribution function $\Phi_{q\bar q}(x,k_t^2)$ 
explicitly and analytically as functions of $m$ and $p_a$.
The calculation of $Z_2$ is novel, particularly for QCD.

The weak point of the present work is its model dependence.
The $\uparrow\downarrow$-model picks out one particular aspect,
namely the strong attraction of the spin-spin interaction in
the singlet channel.
While such description is perhaps viable as indicated by several other
phenomenological models, the above discussion on the model results
might be misleading. 
As listed in Table~\ref{eq:results}, 
the effective quark mass in Eq.(\ref{eq:bsEq})  
is much larger than the bare quark mass in the original QCD-Lagrangian. 
Thus, the solution of reduced wave function in Eq.(\ref{eq:bsEq}) should be 
regarded as the bound state of not bare but dressed quark and antiquark. 
The effective 2-particle bound state is then no longer a compact object. 
In fact, the form factor obtained only with the subset 
($\uparrow\downarrow$) component of light-cone helicity amplitude 
cannot be trusted. 
In multiplying the two light-cone wave functions to compute the 
form factor, the $\uparrow\uparrow$ and $\downarrow\downarrow$ components 
contribute perhaps as significantly as the $\uparrow\downarrow$ and 
$\downarrow\uparrow$ components.
There are also other aspects which must be improved in the future.

One should emphasize however, that the $\uparrow\downarrow$-model
works only with the parameters appearing in the QCD-Lagrangian, 
the quark masses and the coupling constant.
This is kind of the minimum requirement for any effective theory.
inspired by QCD, and an improved solution cannot have 
less information than the momentum spread. 
The model applied serves at least the purpose
to relate $p_a$ to $m$ and $\alpha$ through the (numerical) solution.

\ack
I like to thank both Danny Ashery and Stan Brodsky for having 
commented and improved an early version of the manuscript.
\begin{appendix}
\section{On the evaluation of the integrals}
\label{app:1}
The major labor of this work is in the evaluation of the integrals. 
As a rule, they are straightforward, with one exception: 
Evaluating Eq.(\ref{eq:Prel}) as it stands, even Mathematica drives crazy. 
Therefore some subtle aspects should be mentioned.
From the definition
\begin{equation}
   P(t) = 
   \int\limits_{-1}^{+1} \frac{dz}{2}\ \int\limits_0^\infty du\ %
   \frac{\left[1-z^2\right]\ \left[1+u\right]^{\frac{1}{2}}}
   {\left[1+(z^2-1)t^2+u\right]^2} 
,\end{equation}
follows the particular value $P(0) = {4}/{3}$.
Introducing the new integration variable $x=\sqrt{1-z^2}$
and integrating by parts over $u$, two terms arise in
$P(t) = P_1(t) + P_2(t)$ with 
\begin{equation}
   P_1(t) = 
   \int\limits_{0}^{1} \frac{dx\,x^3}{\sqrt{1-x^2}}\frac{1}{1-t^2x^2}
,\quad\mathrm{and}\quad
   P_2(t) = \int\limits_{0}^{1} \frac{dx\,x^2}{\sqrt{1-x^2}}
   \frac{\mathrm{Arth}(xt)}{t}
.\end{equation}
Note that $P_1(0) =P_2(0) =2/3$.
The first of them is elementary, {\it i.e.}
\begin{equation}
   P_1(t) = \left[
   \frac{\sqrt{1-x^2}}{t^2} - 
   \frac{\mathrm{Arth}\left(\left(t\sqrt{1-x^2}\right)/\sqrt{1-t^2}\right)}
   {t^2\sqrt{1-t^2}}
   \right]_{0}^{1}
,\end{equation}
while $P_2(t)$ has resisted all attempts to integrate it up.
Only when observing the identity
${d\left(tP_2(t)\right)}/{dt} = P_1(t) $
it was possible to proceed with
\begin{equation}
   P_2(t) = \frac{1}{t} \int^t dy\,P_1(y)
.\label{eq:a5}\end{equation}
The (vanishing) integration constant can be determined at the end 
from requiring $P_2(0)=2/3$.
One can insert either the power expansion for
\begin{equation}
   P_1(t) = \frac{2}{3} + \frac{8}{5}\frac{t^2}{3\cdot3} +
   \frac{16}{7}\frac{t^4}{5\cdot5} + \dots,
   \qquad\quad \mathrm{for}\quad \vert t \vert < 1
,\end{equation}
or one can integrate it up in terms of the useful function
\begin{eqnarray}
   F_u(x) = \frac{1}{x}\int^x 
   \frac{dx\,x}{\sin{x}} &=& 1+\sum_{n=1}^\infty
   \frac{2^{2n}-2}{(2n+1)!}\mathrm{B}_n x^{2n}
\\
   &\simeq&
   1 + \frac{x^2}{3\cdot3!} + \frac{7x^3}{3\cdot5\cdot5!}+ \dots
,\end{eqnarray}
where the $\mathrm{B}_n$ are Bernoulli's numbers.
One gets to this form from Eq.(\ref{eq:a5}) by a final
variable transform $y=\sin{x}$, {\it i.e.}
\begin{eqnarray}
   P_2(t) &=& \frac{1}{t^2} + 
   \frac{1}{t}\int^{\mathrm{arcsin}t} \frac{dx\,x}{\sin^3{x}}
\nonumber\\ &=& 
   \frac{1}{2t^2} - 
   \frac{\mathrm{arcsin}t}{2t} \frac{\sqrt{1-t^2}}{t^2}  +
   \frac{\mathrm{arcsin}t}{2t}\,F_u(\mathrm{arcsin}t)
.\end{eqnarray}
In the approximate Eq.(\ref{eq:Prel}), $F_u(\mathrm{arcsin}t)$
was approximated by $1$.
\end{appendix}
\end{document}